\documentstyle[12pt]{article}
\newcommand{\Dslash}{\not \!\! D}
\newcommand{\Aslash}{\not \!\! A}
\newcommand{\kslash}{\not \!\! k}
\newcommand{\delslash}{\not \! \partial}
\begin{document}

\makeatletter
\@addtoreset{equation}{section}
\def\theequation{\thesection.\arabic{equation}}
\makeatother

\begin{flushright}{July 2000\\
UT-901}
\end{flushright}
\vskip 0.5 truecm

\begin{center}
{\large{\bf Aspects of Anomalies in Field Theory\footnote{Talk
given at International Conference on Fundamental Sciences:
Mathematics and Theoretical Physics, Singapore, 13-17 March 2000 
(To be published in the Proceedings, World Scientific, 
Singapore)}}}
\end{center}
\vskip .5 truecm
\centerline{\bf Kazuo Fujikawa}
\vskip .4 truecm
\centerline {\it Department of Physics,University of Tokyo}
\centerline {\it Bunkyo-ku,Tokyo 113,Japan}
\vskip 0.5 truecm

\begin{abstract}
We discuss some formal aspects  of quantum anomalies 
with an emphasis on the regularization of field theory. 
We briefly review how ambiguities  in perturbation theory have 
been resolved by various regularization schemes. 
To single out  the true quantum anomaly among ambiguities,
the combined ideas of  PCAC, soft pion limit and 
renormalizability were essential.
As for the formal treatment of quantum anomalies, we mainly 
discuss the path integral formulation both in continuum and 
lattice theories. In particular,  we discuss in some detail  
the recent development in the treatment of chiral anomalies in 
lattice gauge theory.  
\end{abstract}

\section{Introduction}
The notion of quantum anomalies played key roles in
 various applications of modern field theory. The good accounts 
of these developments and applications are found in 
\cite{adler} \cite{jackiw} \cite{bertlmann}.
Here we discuss some formal aspects  of quantum anomalies with
 an emphasis on the regularization of field theory. 

It is well known that the modern field theory was formulated 
by Tomonaga, Schwinger and Feynman. This modern field theory
with renormalization showed that the principles of quantum 
mechanics are applicable
to quite a wide range of phenomena, covering  all the energies 
presumably up to Planck scales. In the early days of modern 
renormalization theory, however, the treatment
of ultraviolet divergences was a problematic procedure. 
  
\section{Ambiguities in loop diagrams in field theory}

In a letter of  Tomonaga to Oppenheimer,
which reported on a summary of renormalization theory in Japan 
and which was later published in Physical Review\cite{tomonaga0},
Tomonaga mentioned a certain difficulty to preserve the gauge 
invariance of the quadratically divergent vacuum polarization
tensor 
\begin{equation}
\partial^{\mu}\langle T^{\star}V_{\mu}(x)V_{\nu}(y)\rangle =0
\end{equation}
in interaction picture perturbation theory. Here $V_{\mu}(x)$ 
stands for the fermionic vector current to which the photon 
couples. This one-loop diagram for the self-energy of the 
electromagnetic field in fact remained as one of the most 
subtle Feynman diagrams until the developments of modern gauge 
theory starting around early 1970s.

Motivated by this problematic aspect of the vacuum polarization
tensor \cite{miyamoto}, two members of Tomonaga School, Fukuda 
and
 Miyamoto, examined the next simplest diagram,
 namely the triangle diagram corresponding to the process 
$\pi^{0}\rightarrow\gamma\gamma$. They evaluated the 
correlation functions (in the modern notation)
\begin{equation}
\langle T^{\star} A_{\mu}(x)V_{\alpha}(y)V_{\beta}(z)\rangle 
\end{equation}
and 
\begin{equation}
\langle T^{\star}P(x)V_{\alpha}(y)V_{\beta}(z)\rangle.
\end{equation}
Here the axial vector current $A_{\mu}(x)$ and the pseudoscalar
density $P(x)$ naively satisfy the relation
\begin{equation}
\partial^{\mu}A_{\mu}(x)=2miP(x)
\end{equation}
where $m$ stands for the fermion mass. 
But Fukuda and Miyamoto found the violation of gauge 
invariance in the $AVV$ diagram and also a deviation from the 
naive relation (2.4) for the two amplitudes by an explicit 
evaluation in  perturbation theory\cite{fukuda}.

Apparently, Tomonaga was much interested in this discrepancy
and examined the same calculations in (2.2) and (2.3) by using 
the
regularization of Pauli and Villars\cite{pauli}, of which 
preprint was sent to Tomonaga by Pauli. Tomonaga together with 
his associates concluded that the gauge invariance of the $AVV$ 
diagram is maintained by the regularization but the 
above deviation from the naive relation (2.4) was not uniquely 
resolved by the Pauli-Villars regularization\cite{tomonaga}.

Steinberger at Princeton, who learned  the calculation
by Fukuda and Miyamoto through Yukawa ( as is noted in a 
footnote to his paper\cite{steinberger}), also applied the 
brand-new Pauli-Villars regualrization to the calculations of 
various 
decay modes of meson fields. He also arrived at  a 
conclusion\cite{steinberger} similar to that of 
Tomonaga\cite{tomonaga}.
The application of the regularization prescription of Pauli 
and Villars, though it maintained the gauge invariance of the
$AVV$ amplitude, appeared to modify the finite part of $AVV$ 
amplitude and thus the physical life-time of the neutral pion. 

In 1951, Schwinger also examined the entire issue of gauge 
invariance including the vacuum polarization diagram and also the 
above triangle diagrams\cite{schwinger}. He used the so-called 
proper time regularization and partly {\em imposed} the gauge 
invariance. By this way he successfully handled the vacuum 
polarization diagram, but the discrepancy in the triangle 
diagrams remained.

It is interesting that Feynman, unlike Tomonaga and Schwinger,
was apparently not much interested in the above subtle behavior 
of the triangle diagrams. This is presumably due to the fact that 
Feynman was more interested in the totality of Feynman diagrams 
rather than in the naive canonical manipulations such as (2.4).

To summarize the analyses of the vacuum polarization diagram 
and the triangle diagrams around 1950, Nishijima\cite{nishijima}
once mentioned that it was difficult to distinguish 
a subtle difference between the ``ambiguity'' in perturbation 
theory and the true ``anomaly'' at that time. 

\section{Anomalies and regularization in field theory}

In 1969, Bell and Jackiw\cite{bell} at CERN and 
Adler\cite{adler2} at Princeton 
analyzed the triangle diagrams in greater detail. The new 
ingredients in the analysis of Bell and Jackiw were the partially
 conserved axial-vector current (PCAC) and the picture of the 
pion as the Nambu-Goldstone particle\cite{nambu}. 
Bell and Jackiw noticed the inevitable deviation from PCAC if 
one applies  the conventional Pauli-Villars regularization to
the $\sigma$ model which incorporates  PCAC.
They then  showed that one can preserve both of PCAC and gauge
invariance if one uses a modification of  the Gupta's\cite{gupta} 
implementation of the Pauli-Villars regularization, but this 
spoils renormalizability.

 Adler 
on the other hand imposed the gauge invariance on the vector 
vertices and examined what happens with the axial vector vertex.
By this way he showed that the anomaly in the triangle
diagrams  is unavoidable in any sensible Lorentz invariant,
local and renormalizable field theory with vector gauge symmetry.
It has been established  that 
the anomalous behavior of the triangle diagrams is in fact 
{\em necessary} to explain the main decay mode of the neutral 
pion into two photons in the soft pion limit.

It is instructive to see how  the Pauli-Villars
regularization for quark fields works in the analysis of
the neutral pion decay and  anomaly.
If one denotes the axial-vector current and the pseudo-scalar
density associated with the regulator field with mass $M$ by 
$\tilde{A}_{\mu}(x)$ and $\tilde{P}(x)$, respectively, the 
identity (2.4) is replaced by
\begin{equation}
\partial^{\mu}(A_{\mu}(x)+\tilde{A}_{\mu}(x))=2miP(x)
+2Mi\tilde{P}(x).
\end{equation}

For the massive
quarks with $m\neq0$, which breaks chiral symmetry explicitly, 
the pion is also massive. In the soft-pion limit with 
$p_{\mu}\rightarrow 0$ where $p_{\mu}$ is the four-momentum 
carried by the pion, the left-hand side of the above equation
(3.1) goes to zero, and one obtains
\begin{eqnarray}
\lim_{p_{\mu}\sim0}\int dx e^{ipx}2miP(x)&=&
\lim_{p_{\mu}\sim0}-\int dx e^{ipx}2Mi\tilde{P}(x)\nonumber\\
&=&-\lim_{p_{\mu}\sim0}\int dx e^{ipx}
\frac{e^{2}}{16\pi^{2}}\epsilon^{\mu\nu\alpha\beta}F_{\mu\nu}
F_{\alpha\beta}(x).
\end{eqnarray}
The left-hand side of this relation stands for the interpolating
field for the soft-pion, while the right-hand side stands for the 
anomaly in the limit $M\rightarrow\infty$. The two-photon decay 
amplitude of the pion is thus correctly described.

In the Nambu's picture, one starts with the massless quarks
with $m=0$ and thus the ideal Nambu pion is also massless. In 
this case, one obtains from (3.1)
\begin{eqnarray}
\lim_{p_{\mu}\sim0}\int dx e^{ipx}
\partial^{\mu}(A_{\mu}(x)+\tilde{A}_{\mu}(x))
&=&\lim_{p_{\mu}\sim0}\int dx e^{ipx}2Mi\tilde{P}(x)\\  
&=&\lim_{p_{\mu}\sim0}\int dx e^{ipx}
\frac{e^{2}}{16\pi^{2}}\epsilon^{\mu\nu\alpha\beta}F_{\mu\nu}
F_{\alpha\beta}(x).\nonumber
\end{eqnarray}
The left-hand side of this relation gives a non-vanishing matrix
element between the vacuum and the Nambu pion
due to the Nambu-Goldstone theorem, while the right-hand side
stands for the anomaly in the limit $M\rightarrow\infty$. We 
thus obtain the correct pion decay amplitude.

The Pauli-Villars regularization was also successfully used by
Bardeen \cite{bardeen} in the evaluation of the so-called 
non-Abelian gauge anomaly which satisfies the Wess-Zumino 
integrability condition\cite{wess}.

In the revival of field theory, in particular, local gauge theory
starting at the beginning of 1970s, 't Hooft and 
Veltman\cite{'t hooft}
introduced the dimensional regularization. 
With this 
regularization, we can now handle the vacuum polarization tensor 
without any ambiguity. On the other hand, one has a difficulty 
to handle the axial-vector current in this dimensional 
regularization. This regularization is based on the dimensional 
continuation of the algebra
\begin{equation}
\gamma^{\mu}\gamma^{\nu}+\gamma^{\nu}\gamma^{\mu}=2g^{\mu\nu}
\end{equation}
but no consistent  dimensional continuation of $\gamma_{5}$ 
which satisfies
\begin{equation}
\gamma_{5}\gamma^{\mu}+\gamma^{\mu}\gamma_{5}=0. 
\end{equation}
is known. One thus understands the inevitable appearance of the 
triangle (or more generally {\em chiral}) anomaly, while one can 
preserve the vector gauge invariance consistently.

Starting around  the early 1970s, there appeared 
many interesting applications of anomalies, which are nicely 
summarized in refs.\cite{jackiw}\cite{bertlmann}. During this 
period, the 
topological properties of the chiral anomaly were recognized
 and their applications have been discussed. In 
the presence  of the instantons\cite{belavin}, the Atiyah-Singer 
index theorem\cite{atiyah} states the relation
\begin{equation}
n_{+}-n_{-}=\nu
\end{equation}
where $n_{\pm}$ stand for the zero eigenvalue solutions 
of the Euclidean Dirac equation
\begin{equation}
\Dslash\phi_{n}(x)=0
\end{equation}
with the simultaneous chiral eigenvalues
\begin{equation}
\gamma_{5}\phi_{n}(x)=\pm\phi_{n}(x)
\end{equation}
respectively. The $\nu$ in the right-hand of the above equation
(3.6) stands for the Pontryagin number (or instanton number)
\begin{equation}
\nu =\frac{1}{16\pi^{2}}\int d^{4}x tr F\tilde{F}
\end{equation}
The Atiyah-Singer index theorem is rigorously proved in the 
compact manifold such as $S^{4}$. In the context of Euclidean 
flat space-time, the above index relation has been analyzed in 
detail by Jackiw and Rebbi\cite{jackiw2}.

As an interesting and far-reaching application of the topological
properties of the chiral anomaly, 't Hooft\cite{'t hooft2}
pointed out that the proton can decay even in the standard 
Weinberg-Salam theory. In fact the fermion number contains an 
anomaly for general  parity violating Yang-Mills coupling in the 
presence of instantons.

\section{Path integral formulation of anomalies}

As a convenient means to relate the basically classical analysis 
of the Atiyah-Singer index theorem (3.6) and  the quantun field 
theory, a path integral formulation of  quantum  anomaly was 
proposed\cite{fujikawa}.
In particular, the chiral anomaly was identified with the 
non-trivial Jacobian factor under the chiral transformation of 
fermionic variables.

To illustrate the path integral formulation, we
 start with the QCD-type Euclidean path integral
\begin{equation}
\int {\cal D}\bar{\psi}{\cal D}\psi [{\cal D}A_{\mu}]
\exp [ \int \bar{\psi}(i\Dslash-m)\psi d^{4}x + S_{YM}]
\end{equation}
where $\gamma^{\mu}$ matrices are anti-hermitian with 
$\gamma^{\mu}\gamma^{\nu} +
\gamma^{\nu}\gamma^{\mu} = 2g^{\mu\nu}$, and $\gamma_{5}
= - \gamma^{1} \gamma^{2} \gamma^{3} \gamma^{4}$ is hermitian.
The covariant derivative is defined by 
\begin{equation}
\Dslash\equiv \gamma^{\mu}(\partial_{\mu}- igA^{a}_{\mu}T^{a}) 
=\gamma^{\mu}(\partial_{\mu}- igA_{\mu})
\end{equation}
with Yang-Mills generators $T^{a}$. $S_{YM}$ stands for the 
Yang-Mills action and $[{\cal D}A_{\mu}]$ contains a suitable 
gauge fixing.   

To analyze the chiral Jacobian we expand the fermion variables 
\begin{eqnarray}
\psi (x) &=& \sum_{n}a_{n}\varphi_{n}(x)\nonumber\\
\bar{\psi}(x) &=& \sum_{n}\bar{b}_{n}\varphi_{n}^{\dagger}(x)
\end{eqnarray}
in terms of the eigen-functions of hermitian $\Dslash$
\begin{eqnarray}
\Dslash\varphi_{n} (x) &=& \lambda_{n}\varphi_{n} (x)\nonumber\\
\int d^{4}x \varphi_{n}^{\dagger}(x)\varphi_{l} (x) &=& 
\delta_{n,l}
\end{eqnarray}
which diagonalize the fermionic action in (4.1).
The fermionic path integral measure is then written as 
\begin{equation}
{\cal D}\bar{\psi}{\cal D}\psi = \lim_{N\rightarrow \infty}
\prod_{n=1}^{N}
d\bar{b}_{n}da_{n}
\end{equation}
Under an infinitesimal  global chiral transformation
\begin{equation}
\delta\psi = i\alpha\gamma_{5}\psi, \ \ \ \delta\bar{\psi} 
= \bar{\psi}
i\alpha\gamma_{5}
\end{equation}
we  obtain the Jacobian factor
\begin{eqnarray}
J &=& \exp [-2i\alpha  \lim_{N\rightarrow\infty}\sum_{n=1}^{N}
\int d^{4}x \varphi_{n}^{\dagger}(x)\gamma_{5}\varphi_{n} (x)]
\nonumber\\
&=& \exp [-2i\alpha (n_{+} - n_{-})]
\end{eqnarray}
where $n_{\pm}$ stand for the number of eigenfunctions with 
vanishing
eigenvalues and  $\gamma_{5}\varphi_{n}= \pm \varphi_{n}$ in 
(4.4). We here used the relation 
\begin{equation}
\int d^{4}x \varphi_{n}^{\dagger}(x)\gamma_{5}\varphi_{n} (x)=0
\end{equation}
 for $\lambda_{n}\neq 0$ because  
$\Dslash\gamma_{5}\varphi_{n}(x)=-\lambda_{n}\gamma_{5}
\varphi_{n}(x)$. The Atiyah-Singer index theorem 
$n_{+} - n_{-}= \nu$ with  
 Pontryagin index $\nu$ in (3.9), which was  confirmed for one-
instanton sector in $R^{4}$ space by Jackiw and 
Rebbi\cite{jackiw2}, shows that the chiral
Jacobian (4.7) contains the correct information of chiral 
anomaly. 

To extract a local version of the index (i.e., anomaly), we start
 with the expression
\begin{eqnarray}
n_{+} - n_{-}&=& \lim_{N\rightarrow\infty}\sum_{n=1}^{N}
\int d^{4}x \varphi_{n}^{\dagger}(x)\gamma_{5}
f((\lambda_{n})^{2}/M^{2})\varphi_{n} (x)\nonumber\\
&=&\lim_{N\rightarrow\infty}\sum_{n=1}^{N}\int d^{4}x 
\varphi_{n}^{\dagger}(x)\gamma_{5}f(\Dslash^{2}/M^{2})
\varphi_{n} (x)\nonumber\\ 
&\equiv& Tr \gamma_{5}f(\Dslash^{2}/M^{2})
\end{eqnarray}
for {\em any} smooth function $f(x)$ which rapidly goes to zero 
for $x=\infty$ with $f(0)=1$. Since 
$\gamma_{5}f(\Dslash^{2}/M^{2})$ is a well-regularized operator, 
we may now use the plane wave basis of fermionic variables to 
extract an explicit gauge field dependence,  and we  define  a 
{\em local version} of the index  as
\begin{eqnarray}
&&\lim_{M\rightarrow\infty}tr \gamma_{5}f(\Dslash^{2}/M^{2})
\nonumber\\
&\equiv&\lim_{M\rightarrow\infty}\sum_{n=1}^{\infty}
\varphi_{n}^{\dagger}(x)\gamma_{5}f(\Dslash^{2}/M^{2})
\varphi_{n} (x)\nonumber\\
&=& \lim_{M\rightarrow\infty}tr \int \frac{d^{4}k}{(2\pi)^{4}}
e^{-ikx}\gamma_{5}f(\Dslash^{2}/M^{2})e^{ikx}\nonumber\\
&=&\lim_{M\rightarrow\infty}tr \int \frac{d^{4}k}{(2\pi)^{4}}
\gamma_{5}f\{
(ik_{\mu}+ D_{\mu})^{2}/M^{2} - \frac{ig}{4}[\gamma^{\mu}, 
\gamma^{\nu}]F_{\mu\nu}/M^{2}\}\nonumber\\
&=&\lim_{M\rightarrow\infty}tr M^{4}\int 
\frac{d^{4}k}{(2\pi)^{4}}\gamma_{5}f\{
(ik_{\mu}+ D_{\mu}/M)^{2} - \frac{ig}{4}[\gamma^{\mu}, 
\gamma^{\nu}]F_{\mu\nu}/M^{2}\}\nonumber\\
&=&\frac{g^{2}}{32\pi^{2}}tr \epsilon^{\mu\nu\alpha\beta}
F_{\mu\nu}F_{\alpha\beta}
\end{eqnarray}
after a power expansion in $1/M$\cite{fujikawa}. 
We here
used the relation
\begin{equation}
\Dslash^{2} = D_{\mu}D^{\mu} - \frac{ig}{4}
[\gamma^{\mu}, \gamma^{\nu}]F_{\mu\nu}
\end{equation}
and the rescaling of the variable 
$k_{\mu}\rightarrow M k_{\mu}$. 

When one combines (4.9) and (4.10), one establishes the 
Atiyah-Singer index theorem (in $R^{4}$ space)
\begin{equation}
n_{+}-n_{-}=\int d^{4}x\frac{g^{2}}{32\pi^{2}}tr 
\epsilon^{\mu\nu\alpha\beta}
F_{\mu\nu}F_{\alpha\beta}.
\end{equation}
We  note that the local version of the index (anomaly) in 
(4.10) is valid for Abelian theory also.

>From a view point of regularization, we note that the 
 global index (4.9) as well as a local version of the index 
(4.10) are both independent of the regulator  $f(x)$  provided
\cite{fujikawa} 
\begin{equation}
f(0) =1, \ \ \ f(\infty)=0,\ \ \ f^{\prime}(x)x|_{x=0}=f^{\prime}
(x)x|_{x=\infty}=0. 
\end{equation}
Our regulator $f(x)$ imposes gauge invariance, and thus the 
regulator independence  of chiral anomaly  is consitent with the 
analysis of Adler\cite{adler2}, who showed that the chiral 
anomaly is independent of divergence  and perfectly finite and  
well-defined if one imposes gauge invariance on the triangle 
diagram. From the definition of the basic path integral measure 
in (4.5), the present regularization may be called a 
{\em gauge invariant mode cut-off} regularization. 

The Pauli-Villars
regularization is realized in path integral formulation by 
rewriting the fermionic part of the path integral in (4.1) as
\begin{equation}
\int {\cal D}\bar{\psi}{\cal D}\psi{\cal D}\bar{\phi}{\cal D}\phi
\exp [ \int \bar{\psi}(i\Dslash-m)\psi d^{4}x 
+\int \bar{\phi}(i\Dslash-M)\phi d^{4}x ]
\end{equation}
with a {\em bosonic} spinor $\phi$. In this case, the Jacobian 
factors cancel among the contributions from $\psi$ and $\phi$,
and thus the path integral measure becomes invariant under the 
chiral transformation. The (hard) chiral symmetry breaking by the 
mass term $M$ of the regulator field $\phi$ in the limit 
$M\rightarrow\infty$ gives rise to the correct chiral anomaly.
This fact was used in Section 3.

We note that the above path integral formulation works for the 
conformal (or Weyl) anomaly also\cite{fujikawa2}. A particularly
elegant application of the Weyl anomaly was given by Polyakov in
the context of the first quantization of string 
theory\cite{polyakov}.

Among the applications of chiral anomaly, the existence of the 
anomaly even in the Einstein's general coordinate transformation
which was shown by Alvarez-Gaume and Witten\cite{alvarez-gaume} 
enormously influenced our thinking about 
the quantum gravity and string theory. Other interesting 
applications of chiral and Weyl anomalies are found in 
\cite{jackiw}\cite{bertlmann}. 
 
\section{Chiral anomalies in lattice gauge theory}

The lattice theory provides a very powerful  
regularization of path integral. We have recently seen an 
impressive progress in the treatment of fermions and chiral 
anomaly in lattice gauge theory. This progress is based on the 
so-called Ginsparg-Wilson relation\cite{ginsparg} and an explicit 
construction of lattice Dirac operator by 
Neuberger\cite{neuberger}, which is called as the  
overlap Dirac operator for historical reasons. The overlap Dirac 
operator satisfies the Ginsparg-Wilson relation identically and 
it is free of species doubling. 
Hasenfratz, Laliena and Niedermayer proposed an  interesting 
notion of the index theorem in  lattice gauge 
theory\cite{hasenfratz} for the Dirac operator satisfying the 
Ginsparg-Wilson relation, which was in turn used by 
L\"{u}scher\cite{luscher} to identify  a
modified but exact chiral symmetry of lattice fermions. 
By these developments, we can now formulate the chiral anomaly 
for lattice theory in exactly the same manner as the continuum 
path integral. In particular, the chiral anomaly is defined as 
a non-trivial Jacobian in lattice theory also. Some of the 
manipulations become better defined 
in finite lattice theory, though certain aspects of topological 
considerations become more subtle on the discrete lattice space.

We would like to briefly summarize the essence of the lattice 
formulation of chiral anomaly. The lattice fermionic path 
integral is defined by 
\begin{equation}
\int {\cal D}\bar{\psi}{\cal D}\psi
\exp [ \int \bar{\psi}D\psi]
\end{equation}
where the action is defined as a sum over the Euclidean 
hypercubic lattice points.

\subsection{Representation of the Ginsparg-Wilson algebra}

We start with the lattice Dirac operator $D$ which satisfies the 
algebraic relation\cite{fujikawa3}
\begin{equation}
\gamma_{5}(\gamma_{5}D)+(\gamma_{5}D)\gamma_{5}=
2a^{2k+1}(\gamma_{5}D)^{2k+2}
\end{equation}
where $k$ stands for a non-negative integer and $k=0$ corresponds
to the customary Ginsparg-Wilson relation\cite{luscher}. We here 
work on this general form of closed algebra (5.2), since it is 
known that we can construct a generalization of the overlap 
lattice Dirac operator, which is free of species doubling, for 
any value of $k$\cite{fujikawa3}. The parameter $a$ stands for 
the lattice spacing.  When one defines 
\begin{equation}
H\equiv \gamma_{5}aD
\end{equation}
(5.2) is rewritten as 
\begin{equation}
\gamma_{5}H+H\gamma_{5}=2H^{2k+2}
\end{equation}
or equivalently
\begin{equation}
\Gamma_{5}H+\Gamma_{5}H=0
\end{equation}
where we defined 
\begin{equation}
\Gamma_{5}\equiv \gamma_{5}-H^{2k+1}.
\end{equation}
Note that both of $H$ and $\Gamma_{5}$ are hermitian operators;
in Euclidean lattice theory, $D$ itself cannot be hermitian. 

We now discuss a general representation of the algebraic relation
(5.4). The relation (5.4) suggests that if
\begin{equation}
H\phi_{n} = a\lambda_{n}\phi_{n}, \ \ \ (\phi_{n},\phi_{n})=1 
\end{equation}
with a real eigenvalue $a\lambda_{n}$ for the hermitian 
operator $H$, then
\begin{equation}
H(\Gamma_{5}\phi_{n}) = -a\lambda_{n}(\Gamma_{5}\phi_{n}).
\end{equation}
Namely, the eigenvalues $\lambda_{n}$ and $-\lambda_{n}$ are 
always paired if $\lambda_{n}\neq 0$ and 
$(\Gamma_{5}\phi_{n},\Gamma_{5}\phi_{n})\neq 0$.
We also note the relation, which is derived by sandwiching the
relation (5.4) by $\phi_{n}$,
\begin{equation}
(\phi_{n},\gamma_{5}\phi_{n})=(a\lambda_{n})^{2k+1}\ \ \ \ for\ \
 \ \lambda_{n}\neq 0.
\end{equation}
Consequently
\begin{equation}
|(a\lambda_{n})^{2k+1}|= |(\phi_{n},\gamma_{5}\phi_{n})|\leq
||\phi_{n}||||\gamma_{5}\phi_{n}||=1.
\end{equation}
Namely, all the possible eigenvalues are bounded by
\begin{equation}
|\lambda_{n}|\leq \frac{1}{a}.
\end{equation}

We thus  evaluate the norm of $\Gamma_{5}\phi_{n}$
\begin{eqnarray}
(\Gamma_{5}\phi_{n},\Gamma_{5}\phi_{n})
&=& (\phi_{n},(\gamma_{5}-H^{2k+1})(\gamma_{5}-H^{2k+1})\phi_{n})
\nonumber\\ 
&=&(\phi_{n},(1-H^{2k+1}\gamma_{5}-\gamma_{5}H^{2k+1}+H^{2(2k+1)}
)
\phi_{n})\nonumber\\
&=&[1-(a\lambda_{n})^{2(2k+1)}]\nonumber\\
&=&[1-(a\lambda_{n})^{2}][1+(a\lambda_{n})^{2}+...+
(a\lambda_{n})^{4k}]
\end{eqnarray}
where we used (5.9).
By remembering that all the eigenvalues are real, we find that 
$\phi_{n}$ is a ``highest'' state 
\begin{equation}
\Gamma_{5}\phi_{n}=0
\end{equation}
only if 
\begin{equation}
[1-(a\lambda_{n})^{2}]=(1-a\lambda_{n})(1+a\lambda_{n})=0 
\end{equation}
for the Euclidean positive definite inner 
product $(\phi_{n}, \phi_{n})\equiv\sum_{x}\phi_{n}^{\dagger}(x)
\phi_{n}(x)$.

We thus conclude that the states $\phi_{n}$ with 
$\lambda_{n}= \pm \frac{1}{a}$
 are {\em not} paired by the operation $\Gamma_{5}\phi_{n}$ and 
\begin{equation}
\gamma_{5}D\phi_{n}= \pm \frac{1}{a}\phi_{n}, \ \ \ \gamma_{5}
\phi_{n}= \pm \phi_{n}
\end{equation}
respectively. These eigenvalues are in fact the maximum or 
minimum
of the possible eigenvalues of $H/a$ due to (5.11).

As for the vanishing eigenvalues $H\phi_{n}=0$, we find from
(5.4) that $H\gamma_{5}\phi_{n}=0$, namely, 
$H[(1\pm\gamma_{5})/2]\phi_{n}=0$. We thus have 
\begin{equation}
\gamma_{5}D\phi_{n}=0,\ \ \ \gamma_{5}\phi_{n}=\phi_{n} \ \ \ or 
\ \ \ 
\gamma_{5}\phi_{n}=-\phi_{n}.
\end{equation}
\\
To summarize the analyses so far, all the normalizable 
eigenstates $\phi_{n}$ of $\gamma_{5}D=H/a$ are categorized into 
the following 3 classes:\\
\\
(i)\ $n_{\pm}$ (``zero modes''),\\
\begin{equation}
\gamma_{5}D\phi_{n}=0, \ \ \gamma_{5}\phi_{n} = \pm \phi_{n},
\end{equation}
(ii)\ $N_{\pm}$ (``highest states''), \\
\begin{equation}
\gamma_{5}D\phi_{n}= \pm \frac{1}{a}\phi_{n}, \ \ \
\gamma_{5}\phi_{n} = \pm \phi_{n},\ \ \ respectively,
\end{equation}
(iii)``paired states'' with $0 < |\lambda_{n}| < 1/a$,
\begin{equation}
\gamma_{5}D\phi_{n}= \lambda_{n}\phi_{n}, \ \ \ 
\gamma_{5}D(\Gamma_{5}\phi_{n})
= - \lambda_{n}(\Gamma_{5}\phi_{n}).
\end{equation}
Note that $\Gamma_{5}(\Gamma_{5}\phi_{n})\propto \phi_{n}$ for 
$0<|\lambda_{n}|<1/a$.\\

We thus obtain the index relation\cite{hasenfratz}\cite{luscher}
\begin{eqnarray}
Tr\Gamma_{5}&\equiv& \sum_{n}(\phi_{n},\Gamma_{5}\phi_{n})
\nonumber\\
&=&\sum_{ \lambda_{n}=0}(\phi_{n},\Gamma_{5}\phi_{n})
+\sum_{0<|\lambda_{n}|<1/a}(\phi_{n},\Gamma_{5}\phi_{n})
+\sum_{|\lambda_{n}|=1/a}(\phi_{n},\Gamma_{5}\phi_{n})
\nonumber\\
&=&\sum_{\lambda_{n}=0}(\phi_{n},\Gamma_{5}\phi_{n})\nonumber\\
&=&\sum_{\lambda_{n}=0}(\phi_{n},(\gamma_{5}-H^{2k+1})\phi_{n})
\nonumber\\
&=&\sum_{\lambda_{n}=0}(\phi_{n},\gamma_{5}\phi_{n})
\nonumber\\
&=& n_{+} - n_{-} =  index
\end{eqnarray}
where $n_{\pm}$ stand for the number of  normalizable zero modes
with $\gamma_{5}\phi_{n}=\pm\phi_{n}$ in the classification (i) 
above. We here used the fact that 
$\Gamma_{5}\phi_{n}=0$ for the ``highest states'' and that 
$\phi_{n}$ and $\Gamma_{5}\phi_{n}$ are orthogonal to each other
for $0<|\lambda_{n}|<1/a$ since they have eigenvalues 
with opposite signatures.

We note that all the states $\phi_{n}$ with 
$0<|\lambda_{n}|<1/a$, 
which appear pairwise with $\lambda_{n}= \pm |\lambda_{n}|$, 
can be normalized to satisfy the relations
\begin{eqnarray}
\Gamma_{5}\phi_{n}&=&
[1-(a\lambda_{n})^{2(2k+1)}]^{1/2}\phi_{-n},
\nonumber\\
\gamma_{5}\phi_{n}&=&(a\lambda_{n})^{2k+1}\phi_{n}+
[1-(a\lambda_{n})^{2(2k+1)}]^{1/2}\phi_{-n}.
\end{eqnarray}
Here $\phi_{-n}$ stands for the eigenstate with an eigenvalue
opposite to that of $\phi_{n}$.
These states $\phi_{n}$ cannot be the eigenstates of 
$\gamma_{5}$ since 
$|(\phi_{n},\gamma_{5}\phi_{n})|=|(a\lambda_{n})^{2k+1}|<1$. 
  
\subsection{Chiral Jacobian in lattice theory}

The Euclidean path integral for a fermion is defined by
\begin{equation}
\int{\cal D}\bar{\psi}{\cal D}\psi\exp[\int\bar{\psi}D\psi]
\end{equation}
where
\begin{equation}
\int\bar{\psi}D\psi\equiv \sum_{x,y}\bar{\psi}(x)D(x,y)\psi(y)
\end{equation}
and the summation runs over all the points on the lattice.
The relation (5.4) is re-written as 
\begin{equation}
\gamma_{5}\Gamma_{5}\gamma_{5}D+D\Gamma_{5}=0
\end{equation}
and thus the Euclidean action is invariant under the global
 ``chiral'' transformation\cite{luscher}
\begin{eqnarray}
&&\bar{\psi}(x)\rightarrow\bar{\psi}^{\prime}(x)=
\bar{\psi}(x)+i\sum_{z}\bar{\psi}(z)\epsilon\gamma_{5}
\Gamma_{5}(z,x)\gamma_{5}
\nonumber\\
&&\psi(y)\rightarrow\psi^{\prime}(y)=
\psi(y)+i\sum_{w}\epsilon\Gamma_{5}(y,w)\psi(w)
\end{eqnarray}
with an infinitesimal constant parameter $\epsilon$.
Under this transformation, one obtains a Jacobian factor
\begin{equation}
{\cal D}\bar{\psi}^{\prime}{\cal D}\psi^{\prime}=
J{\cal D}\bar{\psi}{\cal D}\psi
\end{equation}
with
\begin{equation}
J=\exp[-2iTr\epsilon\Gamma_{5}]=\exp[-2i\epsilon(n_{+}-n_{-})]
\end{equation}
where we used the index relation (5.20).

We now relate this index appearing in the Jacobian to the 
Pontryagin index of the gauge field in a smooth continuum limit.
We  start with
\begin{equation}
Tr\{\Gamma_{5}f(\frac{(\gamma_{5}D)^{2}}{M^{2}})\}
=Tr\{\Gamma_{5}f(\frac{(H/a)^{2}}{M^{2}})\}
=n_{+} - n_{-}
\end{equation}
Namely, the index is not modified by any  regulator $f(x)$ with 
$f(0)=1$ and $f(x)$ rapidly going to zero for 
$x\rightarrow\infty$, as can be confirmed by using (5.20). This
means that you can use {\em any} suitable $f(x)$ in the 
evaluation of the index by taking advantage of this property.
We then consider a local version of the index
\begin{equation}
tr\{\Gamma_{5}f(\frac{(\gamma_{5}D)^{2}}{M^{2}})\}(x,x)
=tr\{(\gamma_{5}-H^{2k+1})f(\frac{(\gamma_{5}D)^{2}}{M^{2}})\}
(x,x)
\end{equation}
where trace stands for Dirac and Yang-Mills indices; Tr in (5.28) 
includes a sum over the lattice points $x$.  
A local version of the index is not sensitive to the precise 
boundary condition , and  one may take an infinite volume 
limit of the lattice in the above expression. 

We now examine the continuum limit $a\rightarrow 0$ of the above 
local expression (5.29). (This continuum limit corresponds
 to the so-called ``naive'' continuum limit in the context of 
lattice gauge theory.)  We first observe that the term
\begin{equation}
tr\{H^{2k+1}f(\frac{(\gamma_{5}D)^{2}}{M^{2}})\}
\end{equation}
goes to zero in this limit. The large eigenvalues of 
$H=a\gamma_{5}D$ are truncated at the value $\sim aM$ by the 
regulator $f(x)$ which rapidly goes to zero for large $x$. 

We thus examine the small $a$ limit of 
\begin{equation}
tr\{\gamma_{5}f(\frac{(\gamma_{5}D)^{2}}{M^{2}})\}.
\end{equation}
The operator appearing in this expression is well regularized by 
the function $f(x)$ , and we evaluate the above trace by using 
the plane wave basis to extract an explicit gauge field 
dependence.
We consider a square lattice where the momentum is defined in 
the Brillouin zone
\begin{equation}
-\frac{\pi}{2a}\leq k_{\mu} < \frac{3\pi}{2a}.
\end{equation}
We assume that the operator $D$ is free of  species doubling; in 
other words, the operator $D$ blows up rapidly 
($\sim \frac{1}{a}$) for small $a$ in the momentum region 
corresponding to species doublers. The contributions of doublers 
are eliminated by the regulator $f(x)$ in the above expression, 
since
\begin{equation}
tr\{\gamma_{5}f(\frac{(\gamma_{5}D)^{2}}{M^{2}})\}\sim
(\frac{1}{a})^{4}f(\frac{1}{(aM)^{2}})\rightarrow 0
\end{equation}
for $a\rightarrow 0$ if one chooses $f(x)=e^{-x}$, for example. 

We thus examine the above trace in the momentum range of the 
physical species
\begin{equation}
-\frac{\pi}{2a}\leq k_{\mu} < \frac{\pi}{2a}.
\end{equation}
We obtain the limiting $a\rightarrow 0$ 
expression\cite{fujikawa3}
\begin{eqnarray}
&&\lim_{a\rightarrow 0}tr\{\gamma_{5}f(\frac{(\gamma_{5}D)^{2}}
{M^{2}})\}(x,x)\nonumber\\
&=& \lim_{a\rightarrow 0}tr \int_{-\frac{\pi}{2a}}^{\frac{\pi}
{2a}}\frac{d^{4}k}{(2\pi)^{4}}e^{-ikx}\gamma_{5}
f(\frac{(\gamma_{5}D)^{2}}{M^{2}})e^{ikx}\nonumber\\
&=&\lim_{L\rightarrow\infty}\lim_{a\rightarrow 0}tr 
\int_{-L}^{L}\frac{d^{4}k}{(2\pi)^{4}}e^{-ikx}\gamma_{5}
f(\frac{(\gamma_{5}D)^{2}}{M^{2}})e^{ikx}\nonumber\\
&=&\lim_{L\rightarrow\infty}tr \int_{-L}^{L}
\frac{d^{4}k}{(2\pi)^{4}}e^{-ikx}\gamma_{5}
f(\frac{(i\gamma_{5}\Dslash)^{2}}{M^{2}})e^{ikx}\nonumber\\
&\equiv&tr\{\gamma_{5}f(\frac{\Dslash^{2}}{M^{2}})\}
\end{eqnarray}
where  we first take the limit $a\rightarrow 0$ with fixed 
$k_{\mu}$ in $-L\leq k_{\mu} \leq L$, and then take the limit 
$L\rightarrow \infty$. This 
procedure is justified if the integral is well convergent.
 We also {\em assumed} that the 
operator $D$ satisfies  the following relation in the limit 
$a\rightarrow 0$
\begin{eqnarray}
De^{ikx}h(x) &\rightarrow& e^{ikx}(-\kslash+i\delslash
-g\Aslash)h(x)\nonumber\\
&=&i(\delslash+ig\Aslash)(e^{ikx}h(x))
\equiv i\Dslash(e^{ikx}h(x))
\end{eqnarray}
for any {\em fixed} $k_{\mu}$, ($-\frac{\pi}{2a}< k_{\mu}<
\frac{\pi}{2a}$), and a sufficiently smooth function $h(x)$. The 
function $h(x)$ corresponds to the gauge potential in our case, 
which in turn means that the gauge potential $A_{\mu}(x)$
is assumed to vary very little over the distances of the 
elementary lattice spacing. 

The condition (5.36) as well as the absence of species doubling 
are satisfied by the overlap Dirac operator\cite{neuberger} and 
its generalization\cite{fujikawa3}.
The last expression in (5.35) is identical to the continuum 
result (4.10). 
We thus obtain from (5.28) and (5.35) the lattice
index relation
\begin{equation}
n_{+} - n_{-}=Tr\{\Gamma_{5}f(\frac{(\gamma_{5}D)^{2}}{M^{2}})\}
=\int d^{4}x\frac{g^{2}}{32\pi^{2}}tr 
\epsilon^{\mu\nu\alpha\beta}
F_{\mu\nu}F_{\alpha\beta}
\end{equation}
in the continuum limit. A local version of this relation 
leads to the lattice evaluation of chiral anomaly.

\subsection{Fermion number anomaly in chiral lattice gauge 
theory}

As for the chiral fermions on the lattice, the general
algebra (5.2) satisfies the decomposition
\begin{equation}
D=\frac{(1+\gamma_{5})}{2}D\frac{(1-\hat{\gamma}_{5})}{2}
+\frac{(1-\gamma_{5})}{2}D\frac{(1+\hat{\gamma}_{5})}{2}
\end{equation}
with
\begin{equation}
\hat{\gamma}_{5}\equiv\gamma_{5}-2a^{2k+1}(\gamma_{5}D)^{2k+1},
\ \ \ \ \  (\hat{\gamma}_{5})^{2}=1
\end{equation}
by noting 
$\gamma_{5}(\gamma_{5}D)^{2}=(\gamma_{5}D)^{2}\gamma_{5}$ which
can be proved by using (5.2). 

The fermion number non-conservation in the 
chiral theory defined by 
\begin{eqnarray}
&&\int{\cal D}\bar{\psi}{\cal D}\psi
\exp\{\int\bar{\psi}D_{L}\psi\}\nonumber\\
&&\equiv\int{\cal D}\bar{\psi}{\cal D}\psi\exp\{\int\bar{\psi}
\frac{(1+\gamma_{5})}{2}D\frac{(1-\hat{\gamma}_{5})}{2}\psi\}
\end{eqnarray}
follows from the fermion number transformation
\begin{equation}
\psi\rightarrow e^{i\alpha}\psi,\ \ \ \bar{\psi}\rightarrow
\bar{\psi}e^{-i\alpha}.
\end{equation}
If one remembers that the functional spaces of the variables 
$\psi$ and $\bar{\psi}$ are specified by the 
{\em projection operators} $(1-\hat{\gamma}_{5})/2$ and 
$(1+\gamma_{5})/2$, respectively, the Jacobian factor 
for the transformation (5.41) is given by\cite{luscher}
\begin{eqnarray}
J&=&\exp \{i\alpha Tr[\frac{(1+\gamma_{5})}{2}-
\frac{(1-\hat{\gamma}_{5})}{2}]\}\nonumber\\
&=&\exp\{i\alpha Tr{}[\gamma_{5}-(\gamma_{5}aD)^{2k+1}{]}  \} 
=\exp\{i\alpha{[}n_{+}-n_{-}{]}\}
\end{eqnarray}
where the index is defined in (5.20). (To be precise, the 
lattice formulation of chiral non-Abelian gauge theory has not
been established yet, but the chiral $U(1)$ anomaly in (5.42) is 
evaluated without knowing the details of the path integral 
measure of chiral gauge theory.) 

We thus reproduce the well-known fermion number non-conservation
in chiral non-Abelian theory\cite{'t hooft2}.

\section{Conclusion}

We have briefly reviewed some aspects of anomalies with an
emphasis on the regularization of field theory. We discussed
the recent development in lattice thoery in some detail, since 
this subject is relatively new. As for 
more extensive reviews of anomalies, the readers are asked to 
look
at the review articles such as \cite{adler}\cite{jackiw}
\cite{bertlmann}.


\begin{thebibliography}{99}
\bibitem{adler}
S.L. Adler, in {\em Lectures on Elementary Particles and Quantum 
Field Theory}, edited by S. Deser et al. (MIT Press, Cambridge,
Mass., 1970).
\bibitem{jackiw}
S.B. Treiman, R. Jackiw, B. Zumino and E. Witten, {\em Current 
Algebra and Anomalies} (World Scientific, Singapore, 1985).
\bibitem{bertlmann}
R. Bertlmann, {\em Anomalies in Quantum Field Theory} (Oxford
University Press, Oxford, 1996).
\bibitem{tomonaga0}
S. Tomonaga, Phys. Rev. {\bf 74} (1948) 224.
\bibitem{miyamoto}
Y. Miyamoto, Private communication.
\bibitem{fukuda}
H. Fukuda and Y. Miyamoto, Prog. Theor. Phys.{\bf 4} (1949) 49.
\bibitem{pauli}
W. Pauli and F. Villars, Rev. Mod. Phys. {\bf 21} (1949) 434. 
\bibitem{tomonaga}
H. Fukuda, Y. Miyamoto, T. Miyajima, S. Tomonaga, S. Oneda and 
S. Sasaki, Prog. Theor. Phys. {\bf 4} (1949) 477.
\bibitem{steinberger}
J. Steinberger, Phys. Rev. {\bf 76} (1949) 1180.
\bibitem{schwinger}
J. Schwinger, Phys. Rev.{\bf 82} (1951) 664.
\bibitem{nishijima}
K. Nishijima, Bull. Phys. Soc. Jpn.,{\bf 47} (1992) 859.
\bibitem{bell}
J.S. Bell and R. Jackiw, Nuovo Cim. {\bf 60A} (1969) 47.
\bibitem{adler2}
S.L. Adler, Phys. Rev. {\bf 177} (1969) 2426.
\bibitem{nambu}
Y. Nambu and G. Jona-Lasinio, Phys. Rev.{\bf 122} (1961) 345.
\bibitem{gupta}
S.N. Gupta, Proc. Phys. Soc., {\bf A66} (1953) 129.
\bibitem{bardeen}
W.A. Bardeen, Phys. Rev. {\bf 184} (1969) 1848.
\bibitem{wess}
J. Wess and B. Zumino, Phys. Lett. {\bf 37B} (1971) 95.
\bibitem{'t hooft}
G. 't Hooft and M. Veltman, Nucl. Phys. {\bf B44}(1972) 189.\\
C.G. Bollini and J.J. Giambiagi, Nuovo Cim. {\bf B12} (1972) 20.
\bibitem{belavin}
A.A. Belavin,A.M. Polyakov, A.S. Schwartz and Yu.S. Tyupkin,
Phys. Lett. {\bf 59B} (1975) 85. 
\bibitem{atiyah}
M. Atiyah, R. Bott, and V. Patodi, Invent. Math. 
{\bf 19}(1973)279. 
\bibitem{jackiw2}
R. Jackiw and C. Rebbi, Phys. Rev.{\bf D16}(1977)1052. 
\bibitem{'t hooft2}
G. 't Hooft, Phys. Rev. Lett. {\bf 37} (1976) 8.
\bibitem{fujikawa}
K. Fujikawa, Phys. Rev. Lett. {\bf 42} (1979) 1195;Phys. Rev. 
{\bf
 D21} (1980) 2848;{\bf D22} (1980) 1499(E).
\bibitem{fujikawa2}
K. Fujikawa, Phys. Rev. Lett. {\bf 44}(1980)1733 .
\bibitem{polyakov}
A.M. Polyakov, Phys. Lett. {\bf 103B}(1981) 207.
\bibitem{alvarez-gaume}
L. Alvarez-Gaume and E. Witten, Nucl. Phys. {\bf B234} 
(1983) 269.
\bibitem{ginsparg}
P.H. Ginsparg and K.G. Wilson, Phys. Rev. {\bf D25} (1982)2649.
\bibitem{neuberger}
H. Neuberger, Phys. Lett.{\bf B417}(1998)141;{\bf B427}(1998)353.
\bibitem{hasenfratz} 
P. Hasenfratz, V. Laliena and F. Niedermayer, Phys. Lett. 
{\bf B427}(1998)125, and references therein.
\bibitem{luscher}
M. L\"{u}scher, Phys.Lett.{\bf B428} (1998) 342: Nucl.Phys.
{\bf B549} (1999) 295.
\bibitem{fujikawa3}
K. Fujikawa, ``Algebraic generalization of the Ginsparg-Wilson 
relation'', hep-lat/0004012 (to appear in Nucl. Phys. B).
\end{thebibliography}
\end{document}